\documentclass[aps,prl,twocolumn,floatfix]{revtex4}
\usepackage{graphicx}             

\begin{document}
\title{Heavy quasiparticles in the ferromagnetic superconductor
ZrZn$_{2}$}
\author
{S.~J.~C. Yates, G. Santi, S.~M. Hayden, P.~J. Meeson, and S.~B.
Dugdale}
\affiliation{H.~H. Wills Physics Laboratory, University
of Bristol, Tyndall Avenue, Bristol BS8 1TL, United Kingdom}

\date{\today}

\begin{abstract}

We report a study of the de Haas-van Alphen effect in the normal
state of the ferromagnetic superconductor ZrZn$_{2}$.  Our results
are generally consistent with an LMTO band structure calculation
which predicts four exchange-split Fermi surface sheets.
Quasiparticle effective masses are enhanced by a factor of about
4.9 implying a strong coupling to magnetic excitations or phonons.
Our measurements provide insight in to the mechanism for
superconductivity and unusual thermodynamic properties of
ZrZn$_{2}$.

\end{abstract}

\maketitle

The recent reports of the coexistence of ferromagnetism and
superconductivity in UGe$_{2}$ \cite{Saxena00}, ZrZn$_2$
\cite{Pfleiderer01a}, URhGe \cite{aoki01} as well as the discovery
of superconductivity in the non-magnetic hcp phase of Fe under
pressure \cite{shimizu01} have reopened the debate regarding the
relationship of magnetism and superconductivity. Motivated by the
case of liquid $^{3}$He, the search for superconductivity mediated
by spin fluctuations in nearly magnetic materials has a long
history \cite{berk66_fay80}. The observation of superconductivity
in the ferromagnetic phase of ZrZn$_2$ \cite{Pfleiderer01a} (as
well as UGe$_{2}$ and URhGe) together with the extreme sensitivity
of its occurrence to sample purity
\cite{Pfleiderer01a,Mackenzie98}, make it a strong candidate for
magnetically mediated and spin-triplet superconductivity,
as is believed to exist in Sr$_{2}$RuO$_{4}$
\cite{ishida98,Mackenzie98}, although other scenarios are not
excluded \cite{Singh02}.

A detailed knowledge of the electronic structure of ZrZn$_2$ is
crucial to the understanding of both its normal and
superconducting properties.  Thus we report here a detailed study
of the Fermi surface (FS) made using angle-resolved measurements
of the de Haas-van Alphen effect (dHvA). Our results are compared
with an \emph{ab-initio} electronic structure calculation.  We
observe much of calculated Fermi surface which consists of four
exchange-split sheets.  Our study shows that ZrZn$_2$ is
characterized by a large quasiparticle density-of-states (DOS) at
the Fermi energy in the ferromagnetic state, which arises partly
from the band structure and partly from a large mass enhancement.

ZrZn$_2$ has many unusual properties. Below $T_{\mathrm{FM}} =
28.5$~K it becomes ferromagnetic with an ordered moment of $0.17
\mu_{B}$ per formula unit at low temperatures ($T=2$~K). The
intrinsic moment is unsaturated, with an applied magnetic field of
6 Tesla causing a 50\% increase. Compared with other $d$-band
metals, it has an extremely large electronic heat capacity at low
temperatures $C/T$ = 47 mJK$^{-2}$ mol$^{-1}$.  ZrZn$_2$
crystallizes in the C15 cubic Laves structure, with lattice
constant $a = 7.393$~\AA, the Zr atoms forming a tetrahedrally
co-ordinated diamond structure. Its magnetic properties derive
from the Zr 4$d$ orbitals, which have a significant direct
overlap.

Interest in ZrZn$_2$ has been rekindled by the discovery that it
is superconducting at ambient pressure and that both the
superconductivity and ferromagnetism are simultaneously destroyed
by the application of hydrostatic pressure.  For the samples
considered here, the onset temperature for the superconductivity
is $T_\mathrm{SC} \approx 0.6$~K with $B_\mathrm{c2} \approx
0.9$~T.  The most striking feature of ZrZn$_2$, UGe$_{2}$ and
URhGe is perhaps that the \emph{same} electrons are thought to
participate both to superconductivity and ferromagnetism, in
contrast with the situation in other ``magnetic'' superconductors
e.g. borocarbides, RuSr$_2$GdCu$_{2}$O$_{8}$
\cite{Canfield98_Felner97}, where the magnetism and
superconductivity occur in different parts of the unit cell.
Furthermore, ZrZn$_{2}$ is a three-dimensional intermetallic
compound in which the itineracy of the $d$-electrons is almost
unquestionable, whereas some doubts remain about that of the 5$f$
electrons and roles of the strong magnetocrystalline anisotropy
and quasi-two-dimensional electronic structure in UGe$_{2}$ and
URhGe.

The dHvA effect is due to the quantization of the cyclotron motion
of charge carriers and results in a magnetization,
$M_\mathrm{osc}$, which oscillates with magnetic field,
$\mathbf{B}$.  We interpret our results using the
Lifshitz-Kosevich expression \cite{Shoenberg84} in which each
extremal area $A$ of a FS sheet perpendicular to
$\mathbf{\hat{B}}$ gives a contribution to the oscillatory
magnetization,
\begin{equation}
\label{Eq:dHvA}
M_{\mathrm{osc}} \propto B^{\frac{1}{2}} R_D R_T \sin
\left( \frac{2 \pi F}{B} + \gamma \right),
\end{equation}
where $F(\mathbf{\hat{B}}) = (\hbar/ 2\pi e) A(\mathbf{\hat{B}})$.
$R_D$ and $R_T$ are the Dingle and temperature damping factors,
respectively. The temperature dependence of $M_{\mathrm{osc}}$
yields the cyclotron mass, $m^*$, through $R_T = X/\sinh X$ where
$X= 2\pi^2 k_B T/\hbar \omega_c$ and $\omega_c = e B/m^{*}$. This
mass includes all many-body renormalizations and is related to the
corresponding orbit through,
\begin{equation}
\label{Eq:m_c}
m^{*} = \frac{\hbar}{2 \pi}\oint_{\mathrm{FS~orbit}} \frac{dk}{v_F}.
\end{equation}
The quasiparticle lifetime, $\tau$, and the mean-free-path, $l$,
can be determined on each orbit via the field-dependent factor
$R_D= \exp(-\pi m_b/e B \tau)$. Eq.~(\ref{Eq:dHvA}) shows no
``spin-splitting factor'' \cite{Shoenberg84} since in a
ferromagnet, spin-split sheets of the FS are resolved separately.

The dHvA magnetization was measured using a standard low-frequency
field-modulation technique. Data were collected using a top-loading
dilution refrigerator equipped with a 13.5~T magnet reaching temperatures
as low as 30~mK and with a $^3$He system equipped with an 18 T
magnet. High-quality samples ($\rho_{0}$=0.6~$\mu \Omega$cm) were prepared
by directional cooling of a ZrZn$_2$ melt contained in an Y$_2$O$_3$
crucible inside a tantalum bomb \cite{growth_zrzn2}.

\begin{figure*}
\begin{center}
\includegraphics[width=16cm,clip]{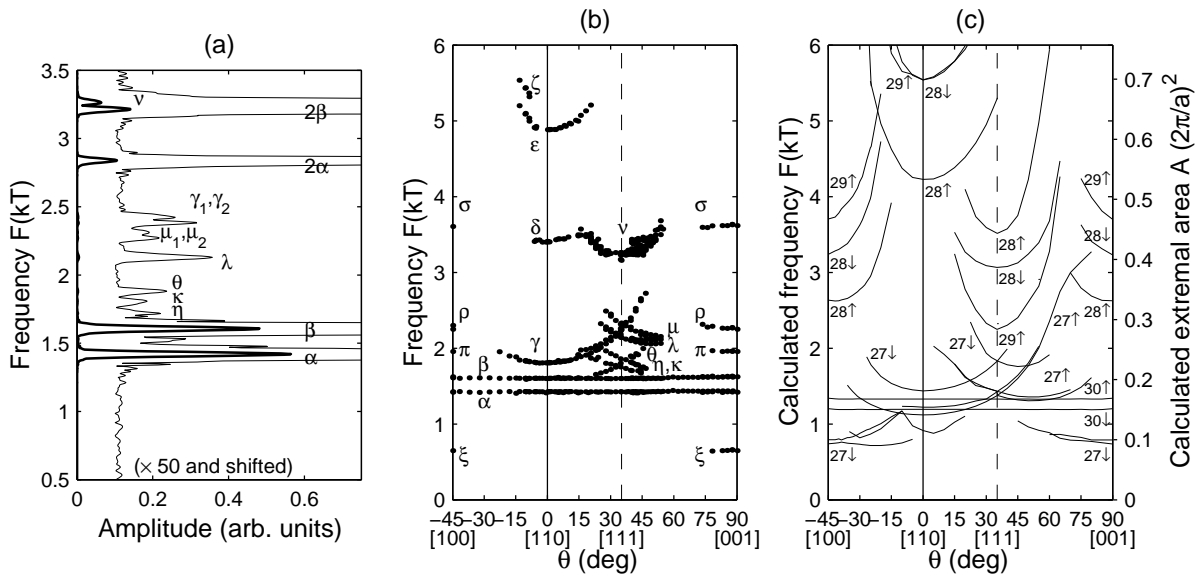}
\end{center}
\caption{
\label{f:rot}
(a) A dHvA spectrum collected for B applied 1.5 deg from
[111] towards [001]. Data was collected in the range $7 < B < 13.4 $~T at
T=25 mK. (b) The angular dependence the dHvA frequencies.  (c) Calculated
dHvA frequencies.}
\end{figure*}

\begin{figure}
\begin{center}
\includegraphics[width=\linewidth,clip]{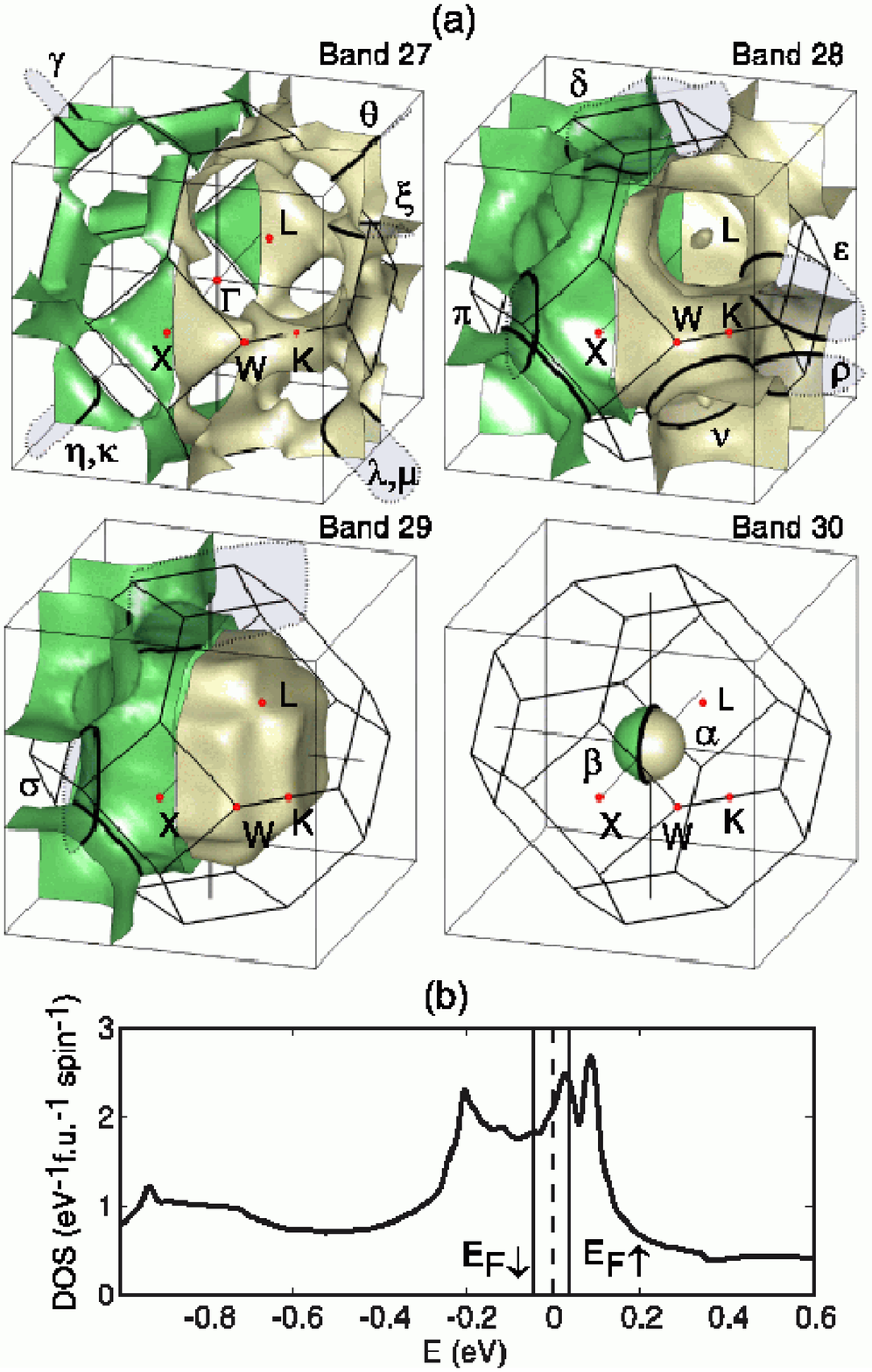}
\end{center}
\caption{ \label{f:fsdos} (a) Calculated Fermi surface of ZrZn$_2$
for the four bands crossing the Fermi level. The left and right
parts of each figure are the majority ($\uparrow$) and
minority ($\downarrow$) spin sheet respectively.  The labeled
solid lines are some of the orbits discussed in the text.  (b) The
calculated DOS for paramagnetic ZrZn$_2$ (see text).  The vertical
solid lines represent the energy shift of the spin-$\uparrow$ and
$\downarrow$ Fermi energies in the ferromagnetic state.}
\end{figure}

We performed angle-resolved dHvA measurements for fields applied
perpendicular to the $[001]$ and $[1\bar{1}0]$ directions.
Fig.~\ref{f:rot}(a) shows a typical Fourier transform of our dHvA data
collected for $\mathbf{\hat{B}}$ close to [111]. The spectrum is dominated
by the low frequencies $F_{\alpha} = 1.42$~kT and $F_{\beta} = 1.60$~kT and
their harmonics.  These frequencies have been previously assigned to an
exchange-split pocket of the Fermi surface centered on the $\Gamma$ point
of the Brillouin zone [see Fig.~\ref{f:fsdos}(a)]. The frequency difference
$F_{\alpha}-F_{\beta}$ is a measure of the ferromagnetic exchange splitting
in ZrZn$_2$.  In addition to $\alpha$ and $\beta$, we also observe a number
of closely spaced frequencies in the range 1.6 -- 2.5~kT ($\eta$, $\kappa$,
$\theta$, $\lambda$, $\mu$, and $\gamma$) and a strong component
($F_{\nu}$) near near 3.2~kT.  Some frequency branches ($\mu$ and $\gamma$)
appear as multiplets.  In all, our study revealed 17 frequency branches,
whose angular dependence is shown in Fig.~\ref{f:rot}(b). Our results are
in good agreement with previous dHvA studies of ZrZn$_2$ in which the
$\alpha$, $\beta$ and $\nu$ branches were observed \cite{dHva_zrzn2}.  In
addition to extremal FS areas, the temperature dependence of the dHvA
amplitude enables the determination of the cyclotron effective mass
corresponding to each orbit (see Table~\ref{t:freq}) from the factor $R_T$
in Eq.~(\ref{Eq:dHvA}).

\begingroup
\squeezetable
\begin{table}
\caption{\label{t:freq} Experimental and calculated dHvA
frequencies and cyclotron masses for the principal high-symmetry
directions in ZrZn$_2$ for $\langle B \rangle = 9.2$T. Branch
assignments refer to Fig.~\ref{f:rot}(b) and the FS orbits are
denoted by orbit center, band index and spin ($\uparrow$ is the
majority).}
\begin{ruledtabular}
\begin{tabular}{ccc|cccc}
  \multicolumn{3}{c|}{Experiment} & \multicolumn{3}{c}{Calculation} \\
branch & $F$ (kT) & $m^*/m_e$ & \multicolumn{2}{c}{FS Orbit} & $F$ (kT)   & $m_{b}/m_e$ \\
\colrule
\multicolumn{7}{c}{[110]} \\
\colrule
$\alpha$                             & 1.425  & 0.81(4) &
$\Gamma_{30,\downarrow}$ & (sphere)  & 1.193  & 0.24  \\
$\beta$                              & 1.600  & 0.95(5) &
$\Gamma_{30,\uparrow}$ & (sphere)    & 1.329  & 0.26  \\
$\gamma$                             & 1.810  & 1.27(3) &
X$_{27,\downarrow}$ & (pillow)       & 1.438  & -0.51 \\
$\delta$                             & 3.400 & 3.7(10) &
X$_{28,\uparrow}$ & (dog bone)       & 4.227  & -2.19 \\
$\epsilon$                           & 4.770  & 3.9(2) &
X$_{28,\downarrow}$ & (dog bone)     & 5.538  & -1.53 \\
$\zeta$                              & 5.300 & 3.5(10) &
                                     &        &       \\
                                     &        &        &
X$_{29,\uparrow}$ & (dog bone)       & 5.541 &  -4.25 \\
\colrule
\multicolumn{7}{c}{[111]}
\\
\colrule
$\alpha$                             & 1.425 & 0.82(5) &
$\Gamma_{30,\downarrow}$ & (sphere)  & 1.193  & 0.24 \\
$\beta$                              & 1.600 & 1.00(4)   &
$\Gamma_{30,\uparrow}$ & (sphere)    & 1.328 & 0.26 \\
$\eta$                               & 1.740 &  3(1)    &
X$_{27,\uparrow}$ & (pillow)         & 1.432 & -1.16  \\
$\theta$                             & 1.809 &  2(1)    &
X$_{27,\uparrow}$ & (pillow)         & 1.376 & -0.64  \\
$\kappa$                             & 1.860 &  3(2)   &
X$_{27,\uparrow}$ & (pillow)         & 1.376 & -0.64  \\
$\lambda_{1}$                        & 2.140 & 1.6(1)  &
X$_{27,\downarrow}$ & (pillow)       & 1.833 & -0.76  \\
$\lambda_{2}$                        & 2.180 &         &
X$_{27,\downarrow}$ & (pillow)       &      &      \\
$\gamma_{1}$                         & 2.240 & 1.9(1)  &
X$_{27,\downarrow}$ & (pillow)       & 1.833 & -0.76 \\
$\gamma_{2}$                         & 2.272 & 3.1(2)  &
X$_{27,\downarrow}$ & (pillow)       &     &        \\
$\mu_{1}$                            & 2.307 & 1.4(3)  &
X$_{27,\downarrow}$ & (pillow)       & 1.833 & -0.76       \\
$\mu_{2}$                            & 2.330 & 2.4(4)  &
X$_{27,\downarrow}$ & (pillow)       &     &        \\
$\nu_1$                              & 3.170 & 2.20(6) &
                & (neck)             &  &       \\
$\nu_2$                              & 3.230 & 1.60(3) &
L$_{28,\downarrow}$ & (neck)         & 3.067 & 0.53   \\
$\nu_3$                              & 3.260 & 1.45(5) &
               & (neck)              &     &        \\
                                     &      &         &
L$_{29,\uparrow}$ & (neck)           & 2.250 & 7.89   \\
                                     &      &         &
L$_{28,\uparrow}$ &  (neck)          & 3.517 & 1.33   \\
\colrule
\multicolumn{7}{c}{[001]} \\
\colrule
$\xi$         & 0.645  & 4(2)    &
W$_{27,\downarrow}$  & (pillow-neck) & 0.767 & 0.68  \\
$\alpha$      & 1.420 & 0.9(1)  &
$\Gamma_{30,\downarrow}$ & (sphere)  &1.201 & 0.24 \\
$\beta$       & 1.610 & 1.1(1)  &
$\Gamma_{30,\uparrow}$& (sphere)     &1.340 & 0.26 \\
$\pi$         & 1.970 & 4(1)    &
W$_{28,\uparrow}$ & (rosette)        & 2.626 & -1.28 \\
$\rho$        & 2.270 & 7(2)   &
W$_{28,\downarrow}$&(rosette)        & 3.232 & -0.94 \\
$\sigma$      & 3.630 & 5(2)    &
W$_{29,\uparrow}$ & (rosette)        & 3.699 & -7.96   \\
\end{tabular}
\end{ruledtabular}
\end{table}
\endgroup

In order to understand our results we performed self-consistent
spin-polarized band structure calculations using the LMTO method
within the local-spin-density approximation (LSDA) \cite{Santi01}.
We obtained a relaxed lattice constant, $a=7.223$~\AA\ and a
magnetic moment, $\mu = 0.18$ $\mu_B$/Zr in excellent agreement
with the measured one (of 0.17 $\mu_B$/Zr \cite{Pfleiderer01a}).
Our calculation agrees well with others in the literature
\cite{bands_zrzn2,Singh02}.  The resulting FS and DOS are shown in
Fig.~\ref{f:fsdos}.  The angular dependence
of the dHvA frequencies [see Fig.~\ref{f:rot}(c)] imposes strong
constraints on the possible topologies of the FS and is therefore
an important tool in verifying the validity of the calculated FS.
The cyclotron band mass for each extremal orbit (see
Table~\ref{t:freq}) is obtained from the relation $m_b =
(\hbar^{2} /2 \pi) \left( \partial A/\partial E \right)_{E_F}$ by
numerical differentiation with a $\Delta E$ of 0.5~mRy.

The \emph{ab-initio} calculated FS appears to reproduce the
essential topology of the observed one.  We assign $\alpha$ and
$\beta$ to the spin-split spherical FS sheet $\Gamma_{30}$ as in
previous studies \cite{dHva_zrzn2}.  The complex of branches
observed near [111] including $\gamma$, $\eta$, $\theta$,
$\kappa$, $\lambda$ and $\mu$ appears to be due to orbits
associated with the X-centered ``pillow'' features of band 27.
Both spin $\uparrow$ and $\downarrow$ FS sheets show this feature,
thus there are pairs of exchange-split branches such as $\theta$
and $\gamma$ and $(\eta,\kappa)$ and $(\lambda,\mu)$. The
$\theta$-branch is probably obscured by the strong $\alpha$ branch
near the [110] direction, only allowing detection near [111].
Another prominent feature of the rotation plot is the $\nu$ branch
which has a minimum frequency for $\mathbf{B}$ $\parallel$ [111].
Some branches, e.g. $\mu$ and $\gamma$ [Fig.~\ref{f:rot}(a)], show
small splittings which could arise from extremely small
corrugations in the FS or a small misalignment of the sample.  The
$\nu$ branch is a signature of the neck around the L point in the
[111] direction and is in excellent quantitative agreement with
the neck for the minority spin ($\downarrow$) of band 28. Although
three sheets of the FS are predicted to have necks along the [111]
directions (orbits L$_{28,\downarrow}$, L$_{28,\uparrow}$,
L$_{29,\downarrow}$) only one neck-like branch is observed
experimentally. The most likely reason is that the higher masses
of the L$_{28,\uparrow}$ and L$_{29,\downarrow}$ orbits (see
Table~\ref{t:freq}) make the signals too weak to be resolved.

The topology of sheet 28 ($\downarrow$ and $\uparrow$ spin) is reminiscent
of the FS of Cu \cite{copper_FS}.
Thus, we expect to see the other orbits seen in Cu. The $\delta$
and $\epsilon$ orbits near $\mathbf{B} \parallel [110]$
(X$_{28,\uparrow}$ and X$_{28,\downarrow}$) correspond to the
``dog bone'' in Cu. We identify the $\pi$ and $\rho$ branches as
the $W_{28,\downarrow}$ and $W_{28,\uparrow}$ calculated orbits
and as the ``four-cornered rosette'' orbits in Cu for fields
$\mathbf{B} \parallel [001]$\cite{copper_FS}. These orbits connect
the [111] necks in neighboring Brillouin zones and are shown in
Fig.~\ref{f:fsdos}(a). The low-frequency $\xi$ branch appears to
be due to the $W_{27,\downarrow}$. Finally, the $\sigma$ branch
appears to be due to either the rosette-type orbit
$W_{29,\uparrow}$ of sheet 29 or an extension of the
X$_{27,\uparrow}$ branch to [001] which may occur with a small
energy shift of band 27.  Our calculation overestimates the size
of the ``dog bone'' and ``rosette'' orbits on sheets 28 and 29.
We note that because all these orbits are related to the necks,
both frequencies would be decreased if these bands were lowered
relative to $E_F$ or if the necks opened out more rapidly along
[111] than the calculation predicts.

We now consider the mass enhancement of the quasiparticles in
ZrZn$_2$. Comparison of the measured and calculated cyclotron
masses shows that the mass enhancements $m^{*}/m_b = 1+\lambda$
are in the range of about 2--5 for $B=9.2$~T. The linear term of
the specific heat $\gamma$ shows a strong field dependence,
$\gamma=35$~mJ~mol$^{-1}$K$^{-2}$ for $B = 9.2$~T and
47~mJ~mol$^{-1}$K$^{-2}$  for $B=0$ \cite{Pfleiderer01b}, thus we
expect $m^{*}$ to be field dependent.  Comparing the measured
$\gamma$ for $B$=9.2~T with $\gamma_\mathrm{calc} = (\pi^2/3)
k_B^2 N_{\mathrm{tot}}(E_F) = 9.6$ mJ~mol$^{-1}$K$^{-2}$
\cite{NF_tot} we find an average mass enhancement $m^{*}/m_b =
3.7$ which is in the center of the range observed by dHvA. If we
make the same calculation for $B=0$, we find $m^{*}/m_b = 4.9$. To
our knowledge, no other alloy of 3$d$-band metals shows a mass
enhancements close to these. Palladium which is is often taken as
an example of a system with strong spin fluctuations has an
enhancement of about two.

\begin{table}
\caption{\label{t:params} Selected measured (calculated)
Fermi surface parameters for ZrZn$_{2}$ \cite{FS_parm}.}
\begin{ruledtabular}
\begin{tabular}{l|ccccc}
branch(orbit)  & $\alpha$($\Gamma_{30,\downarrow}$)  &
$\beta$($\Gamma_{30,\uparrow}$) & $\gamma$(X$_{27,\downarrow}$) &
$\epsilon$(K$_{28,\uparrow}$) \\
\colrule
$F$ (kT) &
1.42(1.19) &
1.60(1.33) &
1.81(1.38) &
4.77(5.54) \\
$\langle k_{F\perp} \rangle$ (\AA$^{-1}$) & 0.21(0.19) &
0.22(0.20) & 0.23(0.20) & 0.38(0.41) \\
$m^{*},m_{b}$ $(m_e)$ &
0.82(0.24) &
0.95(0.26) &
1.27(-0.51) &
3.9(-1.53) \\
$v_F$ (Mms$^{-1}$) &
0.30(0.92) &
0.27(0.90) &
0.21(0.45) &
0.11(0.31)
\\
$\tau (ps)$ &
0.28 &
0.31 &
0.32 &
\\
$l$ (\AA) &
2580 &
2750 &
1500 &
\\
\end{tabular}
\end{ruledtabular}
\end{table}

Table~\ref{t:params} summarizes the important parameters for some
FS orbits in ZrZn$_2$.  The measured Fermi velocity and
cyclotron mass include the effects of electron correlations.  The
quasiparticle lifetime and mean-free-path,$l$, are determined from the Dingle
factor $R_D$.  It is interesting to note that $l
\approx$~1500--2800~\AA\ is considerably larger that the superconducting
coherence length $\xi \approx 270$~\AA\ in this superconducting sample
allowing the possibility of unconventional superconductive pairing
\cite{Mackenzie98}.

\begin{table}
\caption{\label{t:Ni_Fe} Comparison of band parameters (magnetic moment,
$\mu$, exchange-splitting, $\Delta$, DOS at $E_F$ between ZrZn$_2$ and
some $d$-band ferromagnets \cite{Fe_Ni}}
\begin{ruledtabular}
\begin{tabular}{lccc}
    & Fe    & Ni    & ZrZn$_2$  \\
\colrule
$\mu = n_{\uparrow}-n_{\downarrow}$ ($\mu_B$/f.u.)
& 2.12  & 0.56  & 0.16  \\
$\Delta$ (eV)
& 2.2   & 0.6   & 0.07  \\
$N_{\uparrow}(E_F)$(eV$^{-1}$f.u.$^{-1}$ spin$^{-1}$)
& 0.83  & 0.17  & 2.39  \\
$N_{\downarrow}(E_F)$(eV$^{-1}$f.u.$^{-1}$ spin$^{-1}$)
& 0.25  & 1.50  & 1.80  \\
$N_{\mathrm{tot}}(E_F)$(eV$^{-1}$f.u.$^{-1}$)
& 1.08  & 1.67  & 4.19 \\
\end{tabular}
\end{ruledtabular}
\end{table}

It is interesting to compare the properties of ZrZn$_2$ with other $d$-band
ferromagnets. Fig.~\ref{f:fsdos}(b) shows the paramagnetic DOS for
ZrZn$_2$. The Fermi level, $E_F$, lies in a region where the DOS is large
due to the $4d$ Zr band.  Also shown on Fig.~\ref{f:fsdos}(b) are the
relative shifts of $E_F$ for both spins in the ferromagnetic state.  The
small exchange splitting ($\Delta \approx$ 70 meV) of ZrZn$_2$ means
that the Fermi levels for both spins remain in the high DOS region giving
ZrZn$_2$ its unusually large DOS in the ferromagnetic state in contrast
with Fe and Ni (Table~\ref{t:Ni_Fe}). In the ``stronger'' ferromagnets,
$N_{\mathrm{tot}}(\epsilon_F)$ is considerably less in part due to the
larger exchange splitting moving $E_F$ for one of the spin out of the
$d$-band.  A high density DOS at the Fermi energy generally favors
superconductivity so this may be a key factor in understanding the
superconductivity in this material.

In summary, we have used angle-resolved dHvA measurements to characterize
the Fermi surface of the ferromagnetic superconductor ZrZn$_2$. Our
observations are in good agreement with an {\it ab-initio} LSDA band
structure calculation and shed light on some of the unique
properties of ZrZn$_2$.  We find that ZrZn$_2$ has a very large DOS at $E_F$
compared to other $d$-band ferromagnets both because of the
presence of several large sheets of FS and because the
Fermi levels for both majority and minority spins lie in the Zr $d$-band in
the ferromagnetic state.
Further, the fermion quasiparticles are strongly enhanced with an average
mass enhancement of about 4.9. This value is large for materials containing
only $d$-band metallic elements and comparable with that found in the oxide
system Sr$_2$RuO$_4$ \cite{Mackenzie96}.  The presence of the high density
of quasiparticles naturally explain the very large electronic heat capacity
observed in ZrZn$_2$ and also explains the unsaturated magnetic moment (or
large intrinsic magnetic susceptibility) in the ferromagnetic state.
Finally, the high quasiparticle DOS will favor superconductivity.

We are grateful to N.R. Bernhoeft, B.~L. Gy\"orffy, G.G. Lonzarich,
Zs. Major, C. Pfleiderer and J.~B. Staunton for their help with this
work. Support from the EPSRC, the Swiss National Science Foundation (GS)
and the Royal Society (PJM and SBD) is also acknowledged.


\begin{thebibliography}{99}
\bibitem{NF_tot}
We neglect the field dependence of $N_\mathrm{tot}(E_F)$, which we
estimate to be about 10\% for $B \approx 10$~T.
\bibitem{FS_parm}
We calculate the FS parameters as follows: $\left< k_F
\right>=\sqrt{A/\pi}$, $v_F=\hbar\left< k_F \right>/m^{*}$,
$l=v_{F} \tau =\hbar\left< k_F \right> \tau /m_b$.
\end{thebibliography}
\end{document}